\documentclass[12pt,twoside]{article}
\usepackage{fleqn,espcrc1}

\usepackage{graphicx}
\usepackage[figuresright]{rotating}

\newcommand{\s}{$s$}
\newcommand{\ct}{$^{13}$C}
\newcommand{\ctb}{$^{13}$C~}
\newcommand{\nq}{$^{14}$N}
\newcommand{\nqb}{$^{14}$N~}
\newcommand{\cd}{$^{12}$C}
\newcommand{\cdb}{$^{12}$C~}
\newcommand{\msb}{$M_{\odot}$~}
\newcommand{\ms}{$M_{\odot}$}

\newcommand{\nvdb}{$^{22}$Ne~}

\newcommand{\nvdanb}{$^{22}$Ne($\alpha$,n)$^{25}$Mg~}

\newcommand{\ctanb}{$^{13}$C($\alpha$,n)$^{16}$O~}
\newcommand{\nqagb}{$^{14}$N($\alpha$,$\gamma$)$^{18}$F($\beta^+\nu$)$^{18}$O~($\alpha$,$\gamma$)$^{22}$Ne~}

\title{Advances in \s-process models}

\author{M. Lugaro\address{Department of Mathematics, Monash University, 
                       Clayton, Victoria 3800, Australia}%
        \thanks{gratefully acknowledges the support of a IPRS/MGS scholarship,
                    maria.lugaro@maths.monash.edu.au}
        and 
     F. Herwig\address{Universitaet Potsdam, Institut fuer Physik, 
                      Astrophysik, D-14469 Potsdam, Germany}%
        \thanks{present address: Department of Physics \& Astronomy, University of Victoria, B.C., Canada,
                     fherwig@mussel.phys.uvic.ca}}

\begin{document}

% typeset front matter
\maketitle

\begin{abstract}

Within the framework of the current models for the $slow$ neutron capture ($s$) process in Asymptotic Giant
Branch (AGB) stars we explore the uncertainties introduced by the assumptions made on stellar modeling. On
the basis of star models constructed with three different evolutionary codes we generate detailed
neutron-capture nucleosynthesis post-processing models. The main difference among the codes is that one of
them includes an overshooting mechanism. As a result, the neutron fluxes are stronger both during the
interpulse periods and within thermal pulses. The efficiency of the \ctb source is also studied and we find
that a linear relationship exists between the initial \cdb in the intershell and the maximum number of
neutrons produced.

\end{abstract}

\section{INTRODUCTION}

The $s$ process had been suggested in the 1950s to account for the nucleosynthesis of elements heavier than
Fe \cite{burb}. The $s$-process path of neutron capture generally follows the valley of $\beta$ stability,
however, for some values of neutron density and temperature, $branchings$ are open and nuclei outside the
valley of stability can be produced. It can be shown that in equilibrium conditions the $s$-process
distribution follows the rule $\sigma_A N_A = const$, where $\sigma_A$ is the neutron capture cross
section of the isotope $A$ and $N_A$ its abundance in number. The rule is valid locally, far from neutron
magic nuclei which act as bottlenecks of the flux and produce discontinuities in the distribution. An
important parameter is the neutron exposure, $\tau = \int^t_0 N_{\rm n} v_{th} dt$, where $v_{th}$ is the
thermal velocity of the gas. The neutron exposure determines the final distribution of heavy elements;
for higher neutron exposures heavier elements are produced \cite{clayton68}.

The major astrophysical site for the $s$ process has been recognised to be the deep layers of AGB stars,
which are low to intermediate mass stars in a late phase of their evolution. These stars experience
recurrent convective instabilities in the region between the H- and He-burning shells (intershell) as a 
consequence of thermonuclear runaways of the He-burning shell ($thermal$ $pulses$, TP). At the end of a TP,
the convective envelope sinks into the intershell and dredges up material to the surface ($third$ $dredge$
$up$, TDU). A large amount of \nvdb is present in the convective TPs as a product of the chain
\nqagb starting on the abundant \nqb from the H-burning ashes. However, the temperature in the TPs of
low mass stars ($M \leq 4$ \ms) is not high enough to significantly activate the \nvdanb reaction. 
The \ctb neutron source is activated at lower temperatures ($\sim 0.8 \times 10^{8}$ K), however, an
amount of \ctb higher than that present in the H-burning ashes is needed to reproduce the observed
enhancements of heavy elements. To enable the formation of a region rich in \ct, the \ctb $pocket$, some
protons from the envelope must enter the \cd-rich intershell. A favorable location for the occurrence of the
mixing is the sharp H/He discontinuity which is left over after TDU (see Fig. 1), however, no
standard stellar models have found the penetration of protons to occur. Recently, models including
time-dependent overshoot \cite{herwig97}, motivated by hydrodynamical simulations (e.g.\cite{singh98}), and
models with rotation \cite{langer99} have been able to create a proton-rich layer at the top of the
intershell, after the end of the TDU. After less than a few thousand years \ctb and \nqb are formed in the
region. Before the end of the interpulse period all \ctb burns ($\alpha$,n) and the region becomes enriched
in \s-processed material. At the end of the interpulse the pocket is engulfed by the following TP and is thus
mixed with material from the previous TP and the ashes from the H-burning shell. Inside the TP, if the
temperature at the bottom of the shell reaches 2.5 $\times 10^{8}$ K, the \nvdanb reaction is marginally
activated and a second neutron flux can occur. After the TP the $s$-process-rich material from the intershell
is dredged up to the envelope by the following TDU. This cycle is repeated over all TPs with TDU and the
composition of the envelope throughout the AGB phase is changed by TDU and mass loss effects.

Predictions from $s$-process models have to be compared with observational constraints from spectroscopical
observations of chemically peculiar stars (see e.g. \cite{busso98,busso00}) and with laboratory measurements
of isotopic anomalies in silicon carbide (SiC) grains recovered from meteorites, and believed to be
originated in AGB stars \cite{gallino97}. Stellar models, together with Galactic Chemical Evolution models,
should also reproduce the solar distribution of $s$-process elements \cite{trava00}. For an extensive review
on the $s$ process in AGB stars see \cite{busso99}. 

\section{METHOD AND MODELS}

\begin{figure}[t]
\hbox{\hskip -0.2 truein\includegraphics[width=9cm,height=17cm,angle=90]{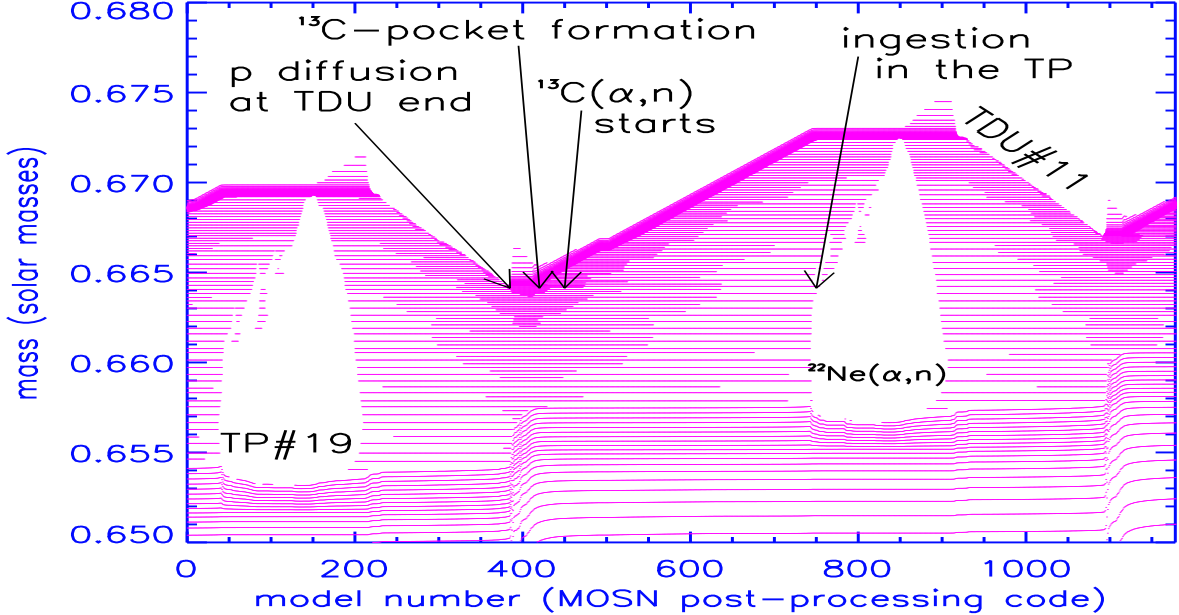}}
\caption{\footnotesize Structure evolution thought two TPs of the 3 \msb star model of solar metallicity as
computed with the MSSSP and represented by the post-processing MOSN code (see \S2). Convective regions are
white and it is possible to recognise the convective zones of the 19$^{\rm th}$ and 20$^{\rm th}$ TPs, as
well as the bottom of the convective envelope. The lines in the radiative regions represent the mass shells,
which evolve in time. The region with the higher mesh density is the H-burning shell. The arrows give a good
representation of the fact that the \ctb pocket is a local phenomenon involving a mass region of the order of
half a tickmark on the y-axis, or less. The bottom of the convective envelope reaches 0.6643 \msb during the
10$^{\rm th}$ TDU and the test exercises described in \S2 have been performed in the region between 0.6638
and 0.6643 \ms.}
%\label{fig:largenenough}
%\label{fig:toosmall}
\end{figure}

The $s$ process has been computed making use of the Torino \s-process code (TOSP, \cite{gallino98}). This
post-processing code calculates neutron captures on all nuclei up to Pb through the AGB phase. Since \ctb burns
in radiative conditions, the \ctb pocket is simulated by specifying independent mass zones of varied extent
and their \ctb amounts. The same \ctb pocket is adopted for all interpulse periods. Reaction rates for
neutron production have been taken from \cite{denker95} for the \ctanb reaction, and from \cite{kappeler94}
for the \nvdanb reaction, excluding the contribution of the elusive resonance at 633 keV. In the temperature
range of interest, from $\sim 2-4 \times 10^8$ K, this rate is within a few percent of the one adopted in
the NACRE compilation \cite{nacre}. Neutron capture rates are updated to the latest estimates \cite{bao}. 

The post-processing models have been generated for a star of 3 \msb and solar metallicity with inputs such as
temperatures, densities, extent of convective zones and composition of the intershell taken from three
different stellar evolutionary codes: the Italian Frascati RAphson Newton Evolutionary Code (FRANEC,
\cite{strani97}), the Australian Mount Stromlo Stellar Structure Program (MSSSP, \cite{frost96}) and the German
EVOL code \cite{herwig00}. Since the codes are independent from one another, they contain several
differences, e.g. with respect to the numerical treatment, opacities and nuclear reaction rates. The most
important difference to be noted here, however, is the treatment of convective instabilities and the
associated mixing. While the FRANEC code uses the Schwarzschild criterion to determine convective boundaries,
the MSSSP code uses a special numerical scheme which may involve mixing of an additional stable mass shell
during structure iterations. The EVOL code employs a time-dependent overshooting mechanism which leads to a
deep penetration of the intershell convection zone into the C/O core during the TP ({\it intershell dredge
up}, IDU), and to the formation of a \ctb pocket over a range of $10^{-6} - 10^{-5}$ \ms. These effects 
depend on the value of the overshoot parameter $f$. The amount of TDU increases when moving from the
FRANEC to the MSSSP to the EVOL code. 

In order to study the neutron flux in the \ctb pocket we have made use of a code specifically designed for
light-element nucleosynthesis in AGB stars: the Monash Stellar Nucleosynthesis (MOSN) code,
which uses the outputs of the MSSSP code. The nuclear network contains 74 light element species, up to
the iron group, and 506 reactions updated as in \cite{lugaro98}. Neutron captures on the missing elements
are modeled by a neutron sink and a fictional particle, $g$, is added to count the number of neutron
captures occurring beyond $^{62}$Ni by an invented decay, with $\lambda$=1 s$^{-1}$: $^{62}$Ni
$\rightarrow$ $^{61}$Ni + g. Since the TOSP code does not includes proton capture reactions, the
MOSN code has been useful in giving a good description of the neutron flux, particularly in regions of the
pocket where the neutron poison reaction \nq(n,p)$^{14}$C and the consequent proton recycle effect is of
importance. We experimented with the MOSN code, artificially introducing some protons below the H/He
discontinuity left by the TDU to form a \ctb pocket. Then we let all the \ctb burn and checked the total
neutron exposure. Since the maximum mass-shell resolution in the region of the pocket at the end of the
interpulse period with the MOSN code is not smaller than $\simeq 5 \times 10^{-4}$ \ms, as shown in Fig. 1,
to perform each test we put a constant amount of protons, rather than a profile, over a mass of $5 \times
10^{-4}$ \ms. 

For the \s-process calculations based on the FRANEC and MSSSP stellar models we used the same \ctb pocket
profile as \cite{gallino98} (see their Fig. 1), which has a total mass of $5 \times 10^{-4}$ \msb and had
been chosen in order to match observational constrains. The \ctb pocket found self-consistently with
the EVOL code and used in our calculation has been computed for the 5$^{\rm th}$ interpulse period with
$f$=0.128 and extents over 1.7 $\times 10^{-5}$ \ms. To compute the resulting neutron flux we performed the
same tests with the MOSN code, however, this time we artificially introduced the \cd, \ctb and \nqb found in 
the pocket with the EVOL code. We therefore specified the \ctb profile in the TOSP code to reproduce the
$\tau$ calculated with MOSN. 

\section{RESULTS AND DISCUSSION}

\subsection{The two neutron bursts}

\begin{figure}[htb]
\begin{minipage}[t]{80mm}
%\framebox[79mm]{\rule[-26mm]{0mm}{52mm}}
\includegraphics[width=79mm,height=79mm]{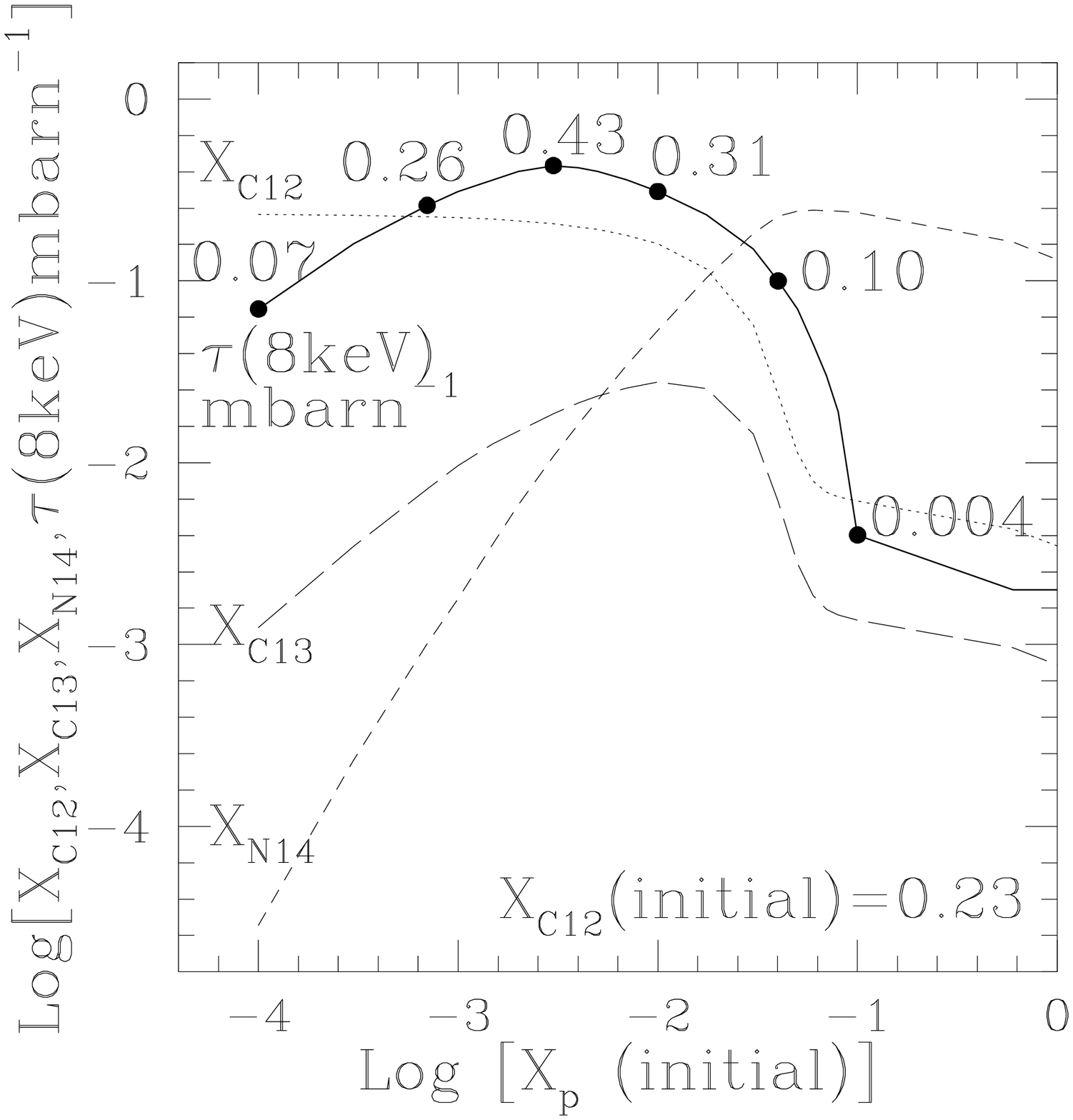}
\caption{\footnotesize Resulting \cd, \ctb and \nqb profiles in the pocket as a function of the initial
number of protons introduced below the H/He discontinuity, after the end of the 10$^{\rm th}$ TDU (see Fig.
1). Also plotted is the total neutron exposure profile as computed at the end of the interpulse period
after all the \ctb has burnt.}
%\label{fig:largenenough}
\end{minipage}
\hspace{\fill}
\begin{minipage}[t]{75mm}
%\framebox[74mm]{\rule[-26mm]{0mm}{52mm}}
\includegraphics[width=79mm,height=79mm]{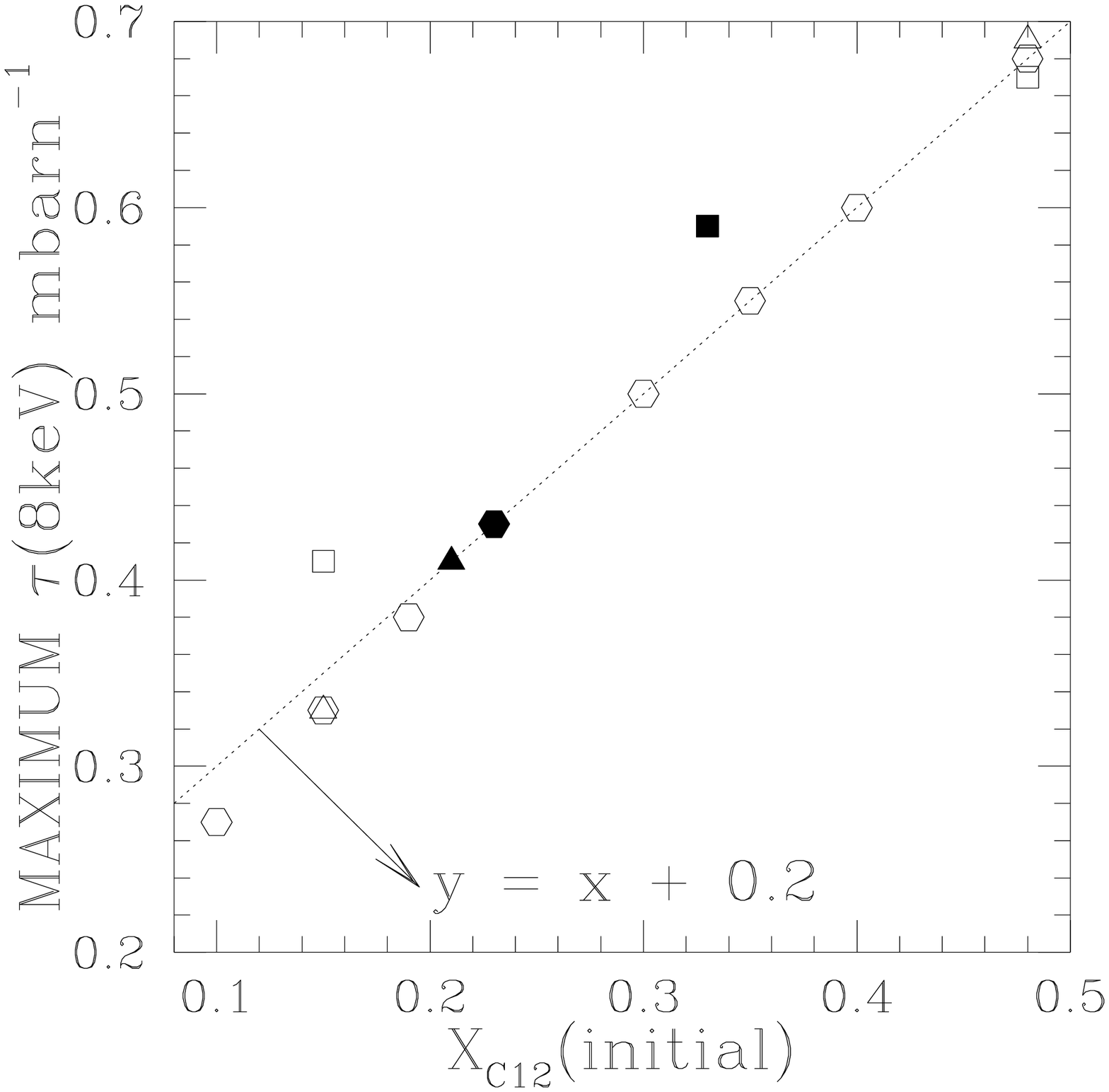}
\caption{\footnotesize Maximum neutron exposure in the pocket as a function of the initial \cd. 
Squares, hexagons and triangles are results obtained for the interpulse periods following the 1$^{\rm st}$, 
10$^{\rm th}$ and 17$^{\rm th}$ TDU respectively. Full symbols indicate the X(\cd) value as computed
by the MSSSP at such interpulses.}
\label{fig:toosmall}
\end{minipage}
\end{figure}

The resulting \cd, \ctb and \nqb profiles in the pocket as a function of the initial X$_p$ are presented in
Fig. 2, as computed with the MOSN code. The total neutron exposure profile obtained when all \ctb has
burnt is also plotted. Since the \nq(n,p)$^{14}$C reaction has a relatively high cross section
\cite{gledenov95}, nuclei of \nqb are a strong neutron poison during the $s$ process. Some of the protons
produced make more \ctb when recaptured by the \cdb (recycle effect), others destroy the \ctb producing more
\nq. As a consequence, the total neutron exposure $\tau$ grows, together with the \ctb profile, up to a
maximum of 0.43 mbarn$^{-1}$, and then decreases when the initial abundance of \nqb nuclei is comparable or
higher than that of \ctb nuclei. This result confirms that the maximum value of $\tau$ in the pocket chosen
by \cite{gallino98} (see their Fig. 3 and 6) is typical for stars of solar metallicity when X(\cd) $\sim$
0.2 in the intershell. We also made some tests changing the initial amount of \cd, and for different
interpulse periods. As shown in Fig. 3 a nearly linear relationship exists between the final maximum value
of $\tau$ in the pocket and the initial \cd, because the latter determines the \ctb and \nqb profiles and
how many protons will be recycled. We note that the \cdb in the intershell varies from pulse to pulse.
This linear relationship suffers from some uncertainties. While the temperature at the H/He interface at
the end of TDU is almost constant ($\sim 10^7$ K) for different interpulse periods, the temperature gradient
towards deeper layers is much steeper for advanced TDU. The temperature at which \ctb and \nqb 
are formed in our calculations, at a mass $\simeq 3 \times 10^{-4}$ \msb below the discontinuity, is
$\sim$ 1.5, 2.2 and 2.5 $\times 10^{7}$ K after the 1$^{\rm st}$, the 10$^{\rm th}$ and the 17$^{\rm th}$ 
TDU respectively. The \cd, \ctb and \nqb profiles can be slightly different and, as shown in Fig. 3, the
$\tau_{max}$ versus \cdb relationship shows some spread. The smaller the extent of the pocket, the less
the change of temperature gradient will effect the results. Future work includes more tests with the MOSN
code, and $s$-process calculations with the TOSP code, using a different $\tau_{max}$ for different
interpulse periods.

As for the EVOL code, because of IDU, the initial mass fraction of \cdb is 0.48, and the resulting
abundance profiles produce a $\tau_{max}$ of 0.88 mbarn$^{-1}$. The discrepancy with the maximum
value of $\tau$ we found for the same initial X(\cd) with the test profiles, computed with the MOSN code,
is possibly due to a slightly lower temperature at which the pocket has been computed, because of the
smaller extent in mass, and mostly to the different choice of the \ct(p,$\gamma$)\nqb reaction rate. In the
MOSN computation the latest rate by \cite{king94} has been adopted, rather than the previous
\cite{cf88}, which is 30\% slower. 

\begin{figure}[t]
\includegraphics[width=16cm,height=9cm]{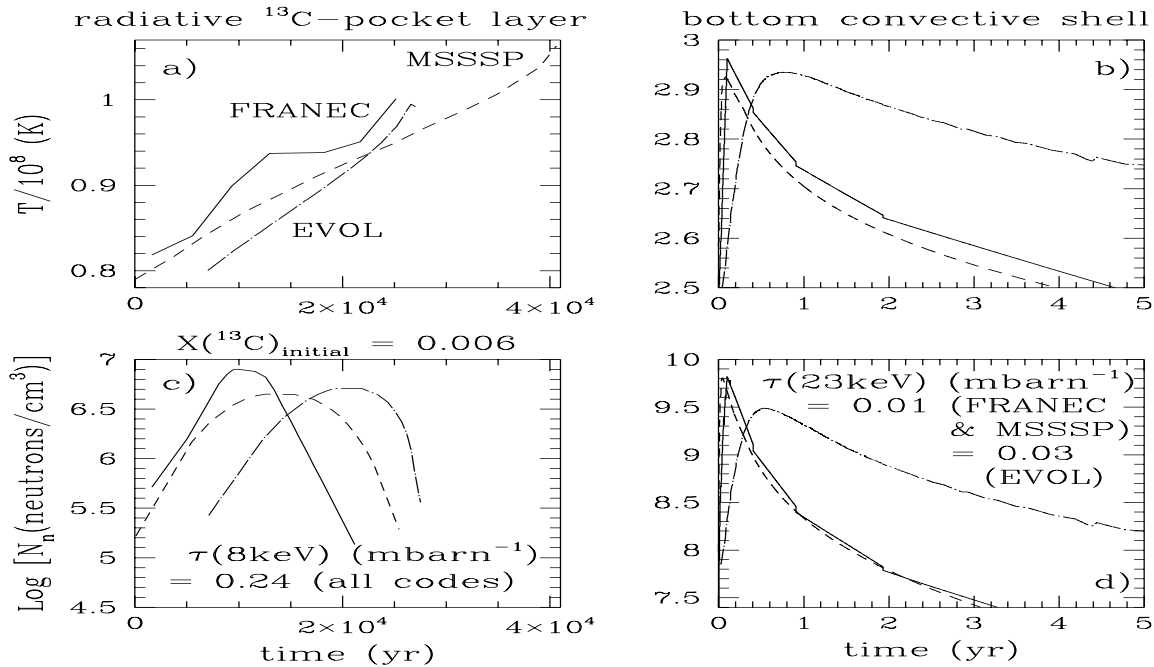}
\caption{\footnotesize The two neutron bursts as described with the different codes: temperature and neutron
density as computed by the TOSP code in one of the zones that forms the \ctb pocket for a typical interpulse
period (panel $a$ and $c$) and at the bottom of a typical TP (panels $b$ and $d$). The zero points in time
represent: in panels $a$ and $c$ the time from the start of the interpulse period (about 10,000 yr)
when $T = 0.8 \times 10^8$ K; in panels $b$ and $d$ the time when $T_{bottom} = 2.5 \times 10^8$ K. Note the
large differences in time scale, total $\tau$ and neutron-density peak between the two neutron bursts.}  
%\label{fig:largenenough}
%\label{fig:toosmall}
\end{figure}

To compare the neutron fluxes in the interpulse period and in the convective pulse, we plot in Fig. 4 the
temperatures and neutron densities for the two neutron bursts, as computed by the TOSP code on the basis of
the three different stellar evolution calculations. This comparison is only of qualitative character because
we did not match the same TP number, core mass and envelope mass simultaneously for the three codes. However,
the main finding is that the features of \ctb burning are fairly comparable in all cases (left panels in Fig.
4). All \ctb burns before the onset of the next TP because the temperature reaches $10^8$ K (panel $a$). This
is a very typical situation, however, it has to be noted and taken into account in future work that this
might not occur during the earliest interpulse periods (see e.g. Fig. 4 of \cite{herwig00}). While the total
neutron exposure is determined only by the initial amount of \ctb in any radiative layer of the pocket, the
maximum neutron density can vary (panel $c$). However, it is so low that no branchings are activated. In the
temperature range of interest the \ctanb rate we use is about 50$-$70\% lower than the one adopted in
\cite{nacre}, however, this should not make much difference in the overall result. If \ctb burns faster the
maximum neutron density will be higher, yet still unimportant. In the TPs, the temperature profile has a
narrow peak (panel $b$), and so has the neutron density (panel $d$). The neutron exposure is unimportant
compared to that in the radiative layers, however, branching points are extremely sensitive to this neutron
flux. Main differences show up between the FRANEC/MSSSP and the EVOL case. In the EVOL code, because of IDU,
the temperature, and consequently the neutron density, stay high for a longer time and branchings are more
active.

\subsection{$s$-process results}

We computed \s-process models with the TOSP code over a total of 25 TPs with TDU for FRANEC, 18 for MSSSP
and 13 for EVOL. In Table 1 the results for the envelope enrichment of \s-process elements are compared
with spectroscopic observations of AGB stars and their relatives. We consider the parameters `ls', which
is the average of light \s-process elements (Y and Zr) and `hs', which is the average of heavy \s-process
elements (Ba, La, Nd and Sm). The ratios of these parameters with Fe, together with the ratios [hs/ls] and
[Pb/Ba] are good indicators of the overall distribution of heavy elements, which is determined by
the main neutron exposure from the \ctb pocket. Predictions are given for the last TP, initial values are
solar (i.e. zero). To compare with measurements in SiC grains we consider three isotopic ratios:
$^{138}$Ba/$^{136}$Ba, $^{88}$Sr/$^{86}$Sr and $^{96}$Zr/$^{94}$Zr. The first two involve $^{138}$Ba and
$^{88}$Sr, which are neutron magic nuclei, and are thus sensitive to the neutron exposure in the \ctb pocket,
while the third ratio involves $^{96}$Zr, which is produced through a branching at $^{95}$Zr, thus it is
sensitive to the neutron density in the TPs. The ranges of prediction are given for the last phase of
the evolution when the condition for SiC grains to form, C/O$>$1, is satisfied in the envelope (last 5 TPs
for the FRANEC, 4 for MSSSP and 7 for EVOL).

\begin{table}[htb]
\caption{Results from the three computations as compared with some observational
constraints.}
\label{table:1}
\newcommand{\cc}[1]{\multicolumn{1}{c}{#1}}
\renewcommand{\tabcolsep}{1.1pc} % enlarge column spacing
\renewcommand{\arraystretch}{1.2} % enlarge line spacing
%\begin{tabular}{@{}lllll}
\begin{tabular}{@{}ccccc}
\hline
parameter & measured & FRANEC & MSSSP & EVOL \\
\hline
$[$ls/Fe$]^a$ & $0. \rightarrow 1.5^b$ & 0.94 & 0.95 & 0.52 \\
$[$hs/Fe$]^a$ & $0. \rightarrow 1.5^b$ & 0.55 & 0.52 & 0.84 \\
$[$hs/ls$]^a$ & $-0.8 \rightarrow 0.5^b$ & $-$0.40 & $-$0.43 & 0.35 \\
$[$Pb/Ba$]^a$ & $-$ & $-$0.52 & $-$0.53 & 0.06 \\
$\delta(^{138}$Ba$/^{136}$Ba)$^c$ & $\sim -250^d$ & $-289 \rightarrow -272$ & $-331 \rightarrow -332$ & $398
\rightarrow 647$ \\
$\delta(^{88}$Sr$/^{86}$Sr)$^c$ & 51$ \pm 103^e$ & $ 208\rightarrow 294$ & $20 \rightarrow 149$ & $158
\rightarrow -293$ \\
$\delta(^{96}$Zr$/^{94}$Zr)$^c$ & $-686 \pm 187^f$ & $-505 \rightarrow -372$ & $-718 \rightarrow -610$ & $46
\rightarrow 1120$ \\
\hline
\end{tabular}\\[2pt]
{\footnotesize $^a$[X/Y]=log[(X/Y)$_{meas}$/(X/Y)$_{\odot}]$.} 
{\footnotesize $^b$Range of values measured in stars of $Z \sim Z_{\odot}$ from \cite{busso98}, typical
2$\sigma$ error $\pm 0.25$.} 
{\footnotesize $^c \delta$(x/y)=[(x/y)$_{meas}$/(x/y)$_{\odot} -$ 1] $\times$ 1000.} 
{\footnotesize $^d$Bulk SiC grains from \cite{prombo93}.} 
{\footnotesize $^e$Average of single grains from \cite{nicolussi98}.} 
{\footnotesize $^f$Average of single grains from \cite{nicolussi97}.}
\end{table}

Since the results from the FRANEC and MSSSP computations, which use the same \ctb pocket, are very similar,
we can conclude that the important uncertainty in \s-process predictions is the \ctb pocket, not the 
intrinsic differences among different codes (with no overshooting mechanism included). In the EVOL
code, the neutron exposure in the pocket is higher and the production of the heavier \s-process elements,
Pb and the magic $^{138}$Ba are favored. The \s-process element enhancements with respect to Fe, of about an
order of magnitude in AGB star envelopes, are reproduced by the models. However, due to the uncertainties
of the spectroscopic measurements on the observational side, and of the \ctb pocket on the theoretical
side, we cannot draw strong conclusions about which of the models match the observations better.
The ratio $^{138}$Ba/$^{136}$Ba measured in bulk SiC grains indicates that, on average, the maximum neutron
exposure in AGB stars should not exceed $\sim$ 0.4 mbarn$^{-1}$. The ratio $^{96}$Zr/$^{94}$Zr is matched when
the neutron density in the TPs is not higher than about $5 \times 10^8$ n/cm$^3$ for a long time, a condition
not satisfied by the EVOL model (see Fig. 4), which produces $^{96}$Zr in a significant amount.

The same MSSSP \s-process calculation was performed with a different \ctb profile, giving more of the total
mass of the pocket to the zone with higher $\tau$ ($\sim 0.4$ mbarn$^{-1}$). Previously, as in
\cite{gallino98}, greater mass was given to the region with lower $\tau$ ($\sim 0.15$ mbarn$^{-1}$). We found
that [ls/Fe] and [hs/Fe] reached 1.27 and 1.0 respectively, while [hs/ls] and [Pb/Ba] kept to $-$0.27 and
$-$0.48. The $\delta$($^{138}$Ba/$^{136}$Ba) did not change much: $-382 \rightarrow -368$. The 
$\delta$($^{88}$Sr/$^{86}$Sr), however, increased to $447 \rightarrow 651$, going outside the
measured range and supporting the profile weighted on the lower $\tau$.

\vspace{2 mm}
{\footnotesize {\bf Acknowledgements.} We wish to thank Roberto Gallino, John Lattanzio,
Maurizio Busso and Onno Pols for discussions and inspiration, Robert Cannon for coding and Brett Hennig
for help with
graphics and proof reading.}


\begin{thebibliography}{9}

\bibitem{burb} E.M. Burbidge et al., Rev. Mod. Phys. 29 (1957) 547.

\bibitem{clayton68} D.D. Clayton, Principles of Stellar Evolution and Nucleosynthesis,
Chicago: Univ. Chicago Press (1968) ch. 7.

\bibitem{herwig97} F. Herwig et al., A\&A 324 (1997) L81.

\bibitem{singh98} H.P. Singh et al., A\&A 340 (1998) 178.

\bibitem{langer99} N. Langer et al., A\&A 346 (1999) L37.

\bibitem{busso98} M. Busso et al., in Nuclei in the Cosmos V, ed. N. Prantzos and S. Harissopulos,
Editions Fronti\`eres Paris (1998) 227. 

\bibitem{busso00} M. Busso et al., these proceedings.

\bibitem{gallino97} R. Gallino et al., in Astrophysical Implications of the
Laboratory Study of Presolar Materials, eds. T. Bernatowicz and E. Zinner, New York: AIP
(1997), 117. 

\bibitem{trava00} C. Travaglio et al., these proceedings.

\bibitem{busso99} M. Busso et al., ARA\&A 37 (1999) 239.

\bibitem{gallino98}R. Gallino et al., ApJ 497 (1998) 388.

\bibitem{denker95} A. Denker et al., in Nuclei in the Cosmos III, ed. M. Busso et al., New York: AIP (1995)
255.

\bibitem{kappeler94} F. K\"appeler et al., ApJ 437 (1994) 396. 

\bibitem{nacre} C. Angulo et al., Nucl. Phys. A656 (1999) 3.

\bibitem{bao}Z.Y. Bao et al., Atom. Data and Nucl. Data Tables 75 (2000) 1. 

\bibitem{strani97} Straniero et al. 1997, ApJ, 478, 332.

\bibitem{frost96} C.A. Frost and J.C. Lattanzio, ApJ 473 (1996) 383.

\bibitem{herwig00} F. Herwig, A\&A 360 (2000) 952.

\bibitem{lugaro98} M. Lugaro, in Nuclei in the Cosmos V, ed. N. Prantzos and S. Harissopulos,
Editions Fronti\`eres Paris (1998) 501. 

\bibitem{gledenov95} Yu.M. Gledenov et al., in Nuclei in the Cosmos III, ed. M. Busso et al., New York: AIP
(1995) 173.
 
\bibitem{king94} J.D. King et al., Nucl. phys. A 567 (1994) 354.

\bibitem{cf88} G.R. Caughlan and W.A. Fowler, Atom. Data Nucl. Data Tables 40 (1988) 283.

\bibitem{prombo93} C.A. Prombo et al., ApJ 410 (1993) 393.

\bibitem{nicolussi98} G.K. Nicolussi et al., Phys. Rev. Lett. 81 (1998) 3583.

\bibitem{nicolussi97} G.K. Nicolussi et al., Science 277 (1997) 1281.

\end{thebibliography}
\end{document}